\documentclass[12pt]{article}

\usepackage{graphicx}
\begin{document}

\begin{center}
{\bf Remarks on Heisenberg$-$Euler-type electrodynamics } \\
\vspace{5mm} S. I. Kruglov
\footnote{E-mail: serguei.krouglov@utoronto.ca}
\underline{}
\vspace{3mm}

\textit{Department of Chemical and Physical Sciences, University of Toronto,\\
3359 Mississauga Road North, Mississauga, Ontario L5L 1C6, Canada} \\
\vspace{5mm}
\end{center}
\begin{abstract}
We consider Heisenberg-Euler-type model of nonlinear electrodynamics with two parameters. Heisenberg-Euler electrodynamics is a particular case of this model. Corrections to Coulomb's law at $r\rightarrow\infty$ are obtained and energy conditions are studied. The total electrostatic energy of charged particles is finite. The charged black hole solution in the framework of nonlinear electrodynamics is investigated. We find the asymptotic of the metric and mass functions at $r\rightarrow\infty$.
Corrections to the Reissner-Nordstr\"{o}m solution are obtained.
\end{abstract}

\section{Introduction}

Classical electrodynamics is modified for strong electromagnetic fields because of self-interaction of photons \cite{Jackson}.
QED becomes nonlinear electrodynamics (NLED) \cite{Heisenberg}, \cite{Schwinger}, \cite{Adler} due to loop corrections and the effect of vacuum birefringence takes place \cite{Rizzo}, \cite{Valle}, \cite{Battesti} \footnote{The BMV and PVLAS experiments provided bounds on the
magnitude of the vacuum birefringence phenomenon.}.
Well-known example of NLED is Born-Infeld (BI) electrodynamics \cite{Born} where an electric field at the origin of particles is finite
as well as the total electrostatic energy.
Problems of singularity and the infinite electromagnetic energy of charges are absent in some NLED models  \cite{Kruglov9}-\cite{Hendi}.
Thus, the attractive feature of NLED is the finiteness of the total electrostatic energy.

The structure of the paper is as follows. In section 2 the NLED model with two parameters is formulated. It is shown that the dual symmetry is  broken in this model. The electric field of point-like charged particles and  corrections to Coulomb's law are obtained. In section 3 we calculate the total electrostatic energy of point-like charged particles which is finite. Energy conditions are investigated.
NLED coupled to gravity is studied and electrically charged black hole is considered in section 4. We
obtain the asymptotic of the metric and mass functions at $r\rightarrow\infty$ which give
corrections to the Reissner-Nordstr\"{o}m (RN) solution. Section 5 is devoted to a conclusion.

We use units with $c=\hbar=1$, $\varepsilon_0=\mu_0=1$ and the metric signature $\eta=\mbox{diag}(-1,1,1,1)$. Greek letters run from $0$ to $3$ and Latin letters run from $1$ to $3$.

\section{A model of nonlinear electrodynamics}

Consider Heisenberg-Euler-type electrodynamics with the Lagrangian density \cite{Kruglov9}
\begin{equation}
{\cal L} = -{\cal F}+\beta{\cal F}^2+\frac{\gamma}{2}{\cal G}^2,
 \label{1}
\end{equation}
where the parameters $\beta$ and $\gamma$ have the dimensions of (length)$^4$, ${\cal F}=(1/4)F_{\mu\nu}F^{\mu\nu}=(\textbf{B}^2-\textbf{E}^2)/2$, ${\cal G}=(1/4)F_{\mu\nu}\tilde{F}^{\mu\nu}=\textbf{E}\cdot \textbf{B}$,
$F_{\mu\nu}=\partial_\mu A_\nu-\partial_\nu A_\mu$ is the field strength tensor, and $\tilde{F}^{\mu\nu}=(1/2)\epsilon^{\mu\nu\alpha\beta}F_{\alpha\beta}$
is the dual tensor. If equalities
\begin{equation}
\beta=\frac{8\alpha^2}{45m_e^4},~~~~\gamma=\frac{28\alpha^2}{45m_e^4},
 \label{2}
\end{equation}
 hold, where $\alpha=e^2/(4\pi)\simeq1/137$ is the fine structure constant, $m_e$ is the electron mass, the model (1) is converted to electrodynamics with quantum corrections (Heisenberg-Euler electrodynamics). In the following we consider parameters, $\beta$ and $\gamma$, as two independent values.
At $\gamma=0$  we arrive at the model considered in \cite{Shabad1}.
The model under consideration, at the weak field limit $\beta {\cal F}\ll 1$, $\gamma{\cal G}\ll 1$, is converted into Maxwell'l electrodynamics, ${\cal L}_M=-{\cal F}$, i.e. the correspondence principle takes place. The nonlinearities of field equations are absent in the weak field limit.
It should be mentioned that a model under investigation is a purely classical model and can be considered only as an effective theory. The nonlinear terms in the Lagrangian do appear from the higher loop corrections in linear QED.

\subsection{Field equations}

From Eq. (1), making use of Euler-Lagrange equations, we find equations of motion
\begin{equation}
\partial_\mu\left[\sqrt{-g}({\cal L}_{\cal F}F^{\mu\nu} +{\cal L}_{\cal G}\tilde{F}^{\mu\nu} )\right]=0,
\label{3}
\end{equation}
where ${\cal L}_{\cal F}=\partial {\cal L}/\partial{\cal F}$, ${\cal L}_{\cal G}=\partial {\cal L}/\partial{\cal G}$, $g= $det$(g_{\mu\nu})$ ($g_{\mu\nu}$ is the metric tensor).
With the aid of Eqs. (1) and (3) one obtains the field equations
\begin{equation}
 \partial_\mu\left[\left(1-2\beta {\cal F}\right)F^{\mu\nu}-\gamma{\cal G}\tilde{F}^{\mu\nu} \right]=0.
\label{4}
\end{equation}
In this section we consider the flat spacetime, and therefore, $\sqrt{-g}=1$.
By virtue of the expression for the electric displacement field $\textbf{D}=\partial{\cal L}/\partial \textbf{E}$, we obtain \cite{Kruglov9}
\begin{equation}
\textbf{D}=\left(1-2\beta {\cal F}\right)\textbf{E}+\gamma {\cal G}\textbf{B}.
\label{5}
\end{equation}
From the relation $\textbf{H}=-\partial{\cal L}/\partial \textbf{B}$ one finds the magnetic field
\begin{equation}
\textbf{H}=\left(1-2\beta {\cal F}\right)\textbf{B}-\gamma{\cal G}\textbf{E}.
\label{6}
\end{equation}
We decompose Eqs. (5) and (6) as  follows \cite{Hehl}:
\begin{equation}
\textbf{D}=\varepsilon \textbf{E}+\nu \textbf{B},~~~~\textbf{H}=\mu^{-1}\textbf{B}-\nu \textbf{E},
\label{7}
\end{equation}
with
\begin{equation}
\varepsilon=1-2\beta {\cal F},~~~~
\mu^{-1}=\varepsilon,~~~~\nu=\gamma {\cal G}.
\label{8}
\end{equation}
The field equations (4), with the help of Eqs. (6) and (7), can be represented as nonlinear Maxwell's equations
\begin{equation}
\nabla\cdot \textbf{D}= 0,~~~~ \frac{\partial\textbf{D}}{\partial
t}-\nabla\times\textbf{H}=0.
\label{9}
\end{equation}
The second pair of nonlinear Maxwell's equations follows from the Bianchi identity, $\partial_\mu \tilde{F}^{\mu\nu}=0$,
\begin{equation}
\nabla\cdot \textbf{B}= 0,~~~~ \frac{\partial\textbf{B}}{\partial
t}+\nabla\times\textbf{E}=0.
\label{10}
\end{equation}
Taking into account Eqs. (5) and (6), we obtain
\begin{equation}
\textbf{D}\cdot\textbf{H}=(\varepsilon^2-\nu^2)\textbf{E}\cdot\textbf{B}+2\varepsilon\nu{\cal F}.
\label{11}
\end{equation}
As $\textbf{D}\cdot\textbf{H}\neq\textbf{E}\cdot\textbf{B}$ \cite{Gibbons} the dual symmetry is violated in the model under consideration. In generalized BI electrodynamics \cite{Krug} and in QED with loop corrections the dual symmetry is also broken but
in classical electrodynamics and in BI electrodynamics the dual symmetry holds.

\subsection{The electric field of point-like charges}

In the presence of point-like charged objects with the electric charge $Q$ the first equation in (9) (in Gaussian units) becomes
\begin{equation}
\nabla\cdot \textbf{D}=4\pi Q\delta(\textbf{r}),
\label{12}
\end{equation}
with the solution
\begin{equation}
 \textbf{D}=\frac{Q}{r^3}\textbf{r}.
\label{13}
\end{equation}
Making use of Eq. (5), at $\textbf{B}=0$,  equation (13) reads
\begin{equation}
E(1+\beta E^2)=\frac{Q}{r^2}.
\label{14}
\end{equation}
It is convenient to explore unitless variables
\begin{equation}
x=\frac{r}{\sqrt{Q}\beta^{1/4}},~~~~y=\sqrt{\beta}E.
\label{15}
\end{equation}
With the aid of Eqs. (15), equation (14) becomes
\begin{equation}
y^3+y-\frac{1}{x^2}=0.
\label{16}
\end{equation}
Cubic equation (16) possesses the solution
\begin{equation}\label{17}
  y(x)=\frac{2}{\sqrt{3}}\sinh\frac{\varphi}{3},~~~\sinh \varphi=\frac{3^{3/2}}{2x^2},~~~\varphi=\ln\left(\frac{3^{3/2}}{2x^2}+
\sqrt{\frac{27}{4x^4}+1}\right).
\end{equation}
The solution to Eq. (14) in another form was obtained in \cite{Shabad1}.
The plot of the function $y(x)$ is given in Fig. 1.
\begin{figure}[h]
\includegraphics[height=3.0in,width=3.0in]{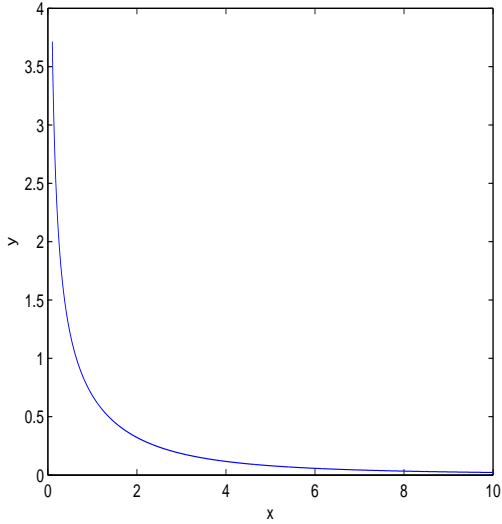}
\caption{\label{fig.1}The plot of the function $y(x)$.}
\end{figure}
The Taylor series of the function $y(x)$ at $x\rightarrow \infty$ ($r\rightarrow \infty$) is
\begin{equation}
y=\frac{1}{x^2}-\frac{1}{x^6}+\frac{3}{x^{10}}-\frac{12}{x^{14}}+{\cal O}(x^{-18}).
\label{18}
\end{equation}
From Eqs. (15) and (18) we obtain the asymptotic electric field at $r\rightarrow\infty$
\begin{equation}
E(r)=\frac{Q}{r^2}-\frac{\beta Q^3}{r^6}+\frac{3\beta^2 Q^5}{r^{10}}-\frac{12\beta^3 Q^7}{r^{14}}+{\cal O}(r^{-18}).
\label{19}
\end{equation}
Eq. (19) gives the corrections to Coulomb's law at $r\rightarrow\infty$ up to ${\cal O}(r^{-18})$.
At $\beta=0$ one comes to Maxwell's electrodynamics and the Coulomb law becomes $E(r)=Q/r^2$.

\section{Energy-momentum tensor and the energy of charged particles}

The symmetrical energy-momentum tensor can be found from the expression
\begin{equation}
T_{\mu\nu}=\left({\cal L}_{\cal F}F_\mu^{~\alpha}+{\cal L}_{\cal G}\tilde{F}_\mu^{~\alpha}\right)F_{\nu\alpha}-g_{\mu\nu}{\cal L}.
\label{20}
\end{equation}
Making use of Eqs. (1) and (20) we obtain the symmetrical energy-momentum tensor
\begin{equation}
T_{\mu\nu}=\left[(2\beta {\cal F}-1)F_\mu^{~\alpha}+
\gamma{\cal G}\tilde{F}_\mu^{~\alpha}\right]F_{\nu\alpha}-g_{\mu\nu}{\cal L}.
\label{21}
\end{equation}
One finds the electric energy density from Eq. (21) (for the case $\textbf{B}=0$)
\begin{equation}
\rho_E=T^{~0}_0=\frac{E^2}{2}\left(1+\frac{3\beta E^2}{2}\right).
\label{22}
\end{equation}
By virtue of Eqs. (15) and (17) we obtain the electric field
\begin{equation}\label{23}
  E(r)=\frac{2}{\sqrt{3\beta}}\sinh\left[\frac{1}{3}\ln\left(\frac{\sqrt{27\beta}Q}{2r^2}+
\sqrt{\frac{27\beta Q^2}{4r^4}+1}\right)\right].
\end{equation}
From Eqs. (22) and (23) one finds
the total electrostatic energy of point-like charges (in Gaussian units)
\begin{equation}
{\cal E}=\int_0^\infty \rho_E r^2dr\simeq \frac{2.47Q^{3/2}}{\beta^{1/4}}.
\label{24}
\end{equation}
We confirm the result obtained in \cite{Shabad1} for the  electrostatic energy of point-like charges within NLED.
As a result, the total electrostatic energy of point-like particles is finite.

\subsection{Energy conditions}

In this subsection we investigate the energy conditions that are of importance for the theory.
If the weak energy condition (WEC) \cite{Hawking} is satisfied then the energy density is positive for any local
observer. WEC is given by
\begin{equation}
\rho\geq 0,~~~\rho+p^m\geq 0 ~~~~ (m=1,~2,~3),
\label{25}
\end{equation}
where $\rho$ is the energy density and $p^m$ are principal pressures, $p^m=-T_m^{~m}$ ($m=1,2,3$, and there is not the summation in the index $m$). Let us consider two cases.

I) $\textbf{B}=0$, $\textbf{E}\neq0$.

The energy density $\rho_E$ is positive that follows from Eq. (22), $\rho_E\geq 0$. From Eq. (21) one obtains
\begin{equation}
p^m_E=-T_m^{~m}=\frac{E^2}{2}\left(1+\frac{\beta E^2}{2}\right)-(1+\beta E^2)E^mE_m~~~~~(m=1,2,3).
\label{26}
\end{equation}
From Eqs. (22) and (26) we find
\begin{equation}
\rho_E+p^m_E=(E^2-E_mE^m)(1+\beta E^2)\geq 0.
\label{27}
\end{equation}
As a result, WEC is satisfied for any value of the electric field.
When the dominant energy condition (DEC) \cite{Hawking} is satisfied then the speed of sound is less than the speed of light. DEC is defined as follows:
\begin{equation}
\rho\geq 0,~~~\rho+p^m\geq 0,~~~\rho-p^m\geq 0~~(m=1,~2,~3).
\label{28}
\end{equation}
DEC includes WEC as it follows from Eqs. (25) and (28).
From Eqs. (22) and (26) one finds
\begin{equation}
\rho_E-p^m_E= E^mE_m(1+\beta E^2)+\frac{\beta E^4}{2}\geq 0.
\label{29}
\end{equation}
Thus, DEC occurs.
The strong energy condition (SEC) \cite{Hawking} defines the acceleration, and it is
\begin{equation}
\rho+\sum_{m=1}^3p^m\geq 0.
\label{30}
\end{equation}
Making use of Eqs. (22) and (26), we obtain
\begin{equation}
\rho_E+\sum_{m=1}^3p^m_E=E^2+\frac{\beta E^4}{2}\geq 0.
\label{31}
\end{equation}
As a result, SEC is satisfied. We find the pressure from the relation
\begin{equation}
p_E= {\cal L}+\frac{E^2}{3}{\cal L}_{\cal F}=\frac{E^2}{6}-\frac{\beta E^4}{12} =\frac{1}{3}\sum_{m=1}^3p_E^m.
\label{32}
\end{equation}
 SEC (30) is equal to $\rho_E+3p_E\geq 0$, and therefore, due to Friedmann's
equation in general relativity, it tells us that electrically charged universe decelerates.
Now we study the second case.

II) $\textbf{E}=0$, $\textbf{B}\neq 0$.

 From Eq. (21)  one obtains
\begin{equation}
\rho_M=\frac{B^2}{2}\left(1- \frac{\beta B^2}{2}\right),
\label{33}
\end{equation}
\begin{equation}
p^m_M=\frac{B^2}{2}\left(1-\frac{3\beta B^2}{2}\right)-B^mB_m(1-\beta B^2).
\label{34}
\end{equation}
By virtue of Eqs. (33) and (34) we find
\begin{equation}
\rho_M+p^m_M=(B^2-B^mB_m)(1-\beta B^2).
\label{35}
\end{equation}
Therefore, WEC is satisfied if $ \beta B^2\leq 1$.
Making use of Eqs. (33) and (34) one obtains
\begin{equation}
\rho_M-p^m_M=\frac{\beta B^4}{2}+B^mB_m(1-\beta B^2).
\label{36}
\end{equation}
Thus, DEC holds if $\beta B^2\leq 1$. We obtain from Eq. (33) and (34)
\begin{equation}
\rho_M+\sum_{m=1}^3p^m_M= B^2\left(1-\frac{3\beta B^2}{2}\right).
\label{37}
\end{equation}
 SEC is satisfied at $B^2\leq 2/(3\beta)$.
The pressure, for our case, is given by
\begin{equation}
p_M={\cal L}-\frac{2B^2}{3}{\cal L}_{\cal F}=\frac{B^2}{6}-\frac{5\beta B^4}{12}=\frac{1}{3}\sum_{m=1}^3p_M^m.
\label{38}
\end{equation}
We make a conclusion, due to Friedmann's equation in general relativity, that magnetized universe decelerates when the average magnetic field squired obeys the inequality $B^2\leq 2/(3\beta)$ and accelerates at $B^2\geq 2/(3\beta)$.

\section{Electrically charged black hole}

Let us consider electrically charged black hole ($\textbf{B}=0$). The action of NLED in general relativity is given by
\begin{equation}
I=\int d^4x\sqrt{-g}\left(\frac{1}{2\kappa^2}R+ {\cal L}\right),
\label{39}
\end{equation}
where $\kappa^2=8\pi G\equiv M_{Pl}^{-2}$, $G$ is Newton's constant, $R$ is the Ricci scalar, and $M_{Pl}$ is the reduced Planck mass.
Varying action (39) with respect to the metric and electromagnetic potentials we obtain the Einstein equation and the equation of motion for electromagnetic fields
\begin{equation}
R_{\mu\nu}-\frac{1}{2}g_{\mu\nu}R=-\kappa^2T_{\mu\nu},
\label{40}
\end{equation}
\begin{equation}
\partial_\mu\left[(\sqrt{-g})F^{\mu\nu}\left(\beta E^2+1\right)\right]=0.
\label{41}
\end{equation}
The line element in the case of the spherical symmetry is given by
\begin{equation}
ds^2=-f(r)dt^2+\frac{1}{f(r)}dr^2+r^2(d\vartheta^2+\sin^2\vartheta d\phi^2).
\label{42}
\end{equation}
Then Eq. (41) becomes
\begin{equation}
\partial_r\left[r^2E(\beta E^2+1)\right]=0,
\label{43}
\end{equation}
with the solution (23).
The metric function is defined by \cite{Bronnikov}
\begin{equation}
f(r)=1-\frac{2GM(r)}{r},
\label{44}
\end{equation}
with the mass function
\begin{equation}
M(r)=\int_0^r\rho_E(r)r^2dr=m_E-\int^\infty_r\rho_E(r)r^2dr.
\label{45}
\end{equation}
Here we imply that the mass of the black hole $m_E$ possesses the electromagnetic nature, i.e. $m_E\equiv {\cal E}=\int_0^\infty\rho(r)r^2dr$ and the electrostatic energy of the black hole is given by Eq. (24).
Making use of Eqs. (19) and (45) we obtain the asymptotic mass function at $r\rightarrow\infty$
\begin{equation}
M(r)=m_E-\frac{Q^2}{2r}+\frac{\beta Q^4}{20r^5}-\frac{\beta^2 Q^6}{18r^{9}}+{\cal O}(r^{-12}).
\label{46}
\end{equation}
Taking into account Eqs. (44) and (46) we obtain the metric function at $r\rightarrow\infty$
\begin{equation}
f(r)=1-\frac{2Gm_E}{r}+\frac{G Q^{2}}{r^2}-\frac{\beta G Q^4}{10r^6}+\frac{\beta^2G Q^6}{9r^{10}}+{\cal O}(r^{-13}).
\label{47}
\end{equation}
Equation (47) has the RN form with corrections in the order of ${\cal O}(r^{-6})$.
At $r\rightarrow \infty$ we have $f(\infty)=1$ and the spacetime becomes flat. Putting $\beta=0$ in Eq. (47)  we arrive at
Maxwell's electrodynamics and  (47) becomes the RN solution.

We obtain the Ricci scalar from Eqs. (21) and (40)
\begin{equation}
R=\kappa^2 T_\mu^{~\mu}=\kappa^2\beta E^4.
\label{48}
\end{equation}
At $r\rightarrow \infty$ the electric field approaches to zero (see Eq. (19)), and therefore, the Ricci scalar goes to zero, $R\rightarrow 0$. As a result, spacetime becomes flat at $r\rightarrow \infty$.

 \section{Conclusion}

We have considered the NLED model with two independent parameters $\beta$ and $\gamma$. At the values (2) one comes to QED with one loop corrections (Heisenberg-Euler electrodynamics). For weak fields the self-interaction of photons is negligible and the model becomes Maxwell's electrodynamics. The phenomenon of vacuum birefringence holds if $\gamma\neq 2\beta$ \cite{Kruglov9}. The dual symmetry is broken in the model. Energy conditions were investigated and we demonstrated that WEC, DEC and SEC are satisfied for the case $\textbf{B}=0$, $\textbf{E}\neq 0$.
We obtained the corrections to Coulomb's law at $r\rightarrow\infty$ that are in the order of ${\cal O}(r^{-6})$. The total electrostatic energy of charged particles is finite. NLED coupled to the gravitational field was considered and electrically charged black hole was studied. We found the asymptotic of the metric and mass functions at $r\rightarrow\infty$ which give corrections to the RN solution.

\end{document}